# Spatial Distribution of the Starbursts in Post-Starburst Coma Cluster Galaxies

Nelson Caldwell

F.L. Whipple Observatory, Smithsonian Institution, Box 97, Amado AZ 85645
Electronic mail: caldwell@flwo99.sao.arizona.edu

James A. Rose

Department of Physics and Astronomy, University of North Carolina, Chapel Hill, NC 27599
Electronic mail: jim@wrath.physics.unc.edu

Marijn Franx

Kapteyn Laboratory, PO Box 800, 9700 AV Groningen, The Netherlands
Electronic mail: franx@cosmos.astro.rug.nl

Andrew Leonardi

Department of Physics and Astronomy, University of North Carolina, Chapel Hill, NC 27599
Electronic mail: leonardi@augustus.physics.unc.edu

## ABSTRACT

We present long slit spectra and multi-color CCD images which demonstrate that the strong star formation episodes that occurred in the post-starburst galaxies in the Coma Cluster and two field galaxies were not restricted to the central regions of the galaxies. Rather, the remnant young stars from the starbursts are found to be distributed over a large radius, though changes in the relative strength of the old and young components with radius are evident in a few cases. The Coma galaxies are shown to have exponential disk profiles, and the spectra provide further kinematical evidence that the galaxies are rotating systems, indicating that the galaxies are not ellipticals. The starburst material also appears to be distributed in disks. Such information places constraints on models for the starbursts that involve mergers.

Population models are discussed in a companion paper, which show that the post-starburst spectra in these particular galaxies can not be reproduced by normal spiral-like star formation which simply ceased, but rather are due to a burst of star formation which ceased $\sim$1 Gyr ago and which presently account for $\sim$60% of the light at 4000Å.

Curiously, an on-going starburst galaxy, previously identified in Caldwell et al. (1993) and also observed here, has its vigorous star formation isolated to its bulge region. The associated ionized gas shows peculiar kinematics, when compared to the gas located in the disk. Of four other Coma early-type galaxies with ionized gas studied here, two are shown to have emission spectra characteristic of active galactic nuclei, while two have spectra showing a mixture of H II region and AGN spectra.

*Subject headings:* galaxies: post-starburst — galaxy clusters: Coma

## 1. INTRODUCTION

Unusual star formation histories in galaxies in clusters and the field have been widely observed at redshifts greater than 0.2 (Butcher & Oemler 1978, 1984, Dressler & Gunn 1983, Couch & Sharples 1987), and more recently in the nearby Coma cluster (Caldwell et al. 1993). In some cases, the galaxies are caught in the act of making stars at a rate much higher than would be the case for a normal spiral galaxy (Lavery & Henry 1986). The majority of cases are those in which a starburst occurred but which ended soon thereafter, after about 1 Gyr. These objects are referred to as post-starburst or E+A galaxies (Gunn 1988), the latter a description of the spectra which appear to have an A-star component added in to an old elliptical like component (Franx, 1993, has suggested that the term "K+A" might be less of a mixed metaphor). Suggestions about what triggers these bursts range from mergers with nearly equal mass neighbors (Lavery & Henry 1994, Barnes & Hernquist 1991) to galaxies running into hot cluster gas, which ignites a brief but energetic episode of star formation (Gunn 1989).

Key to distinguishing among such musings is the physical location of the starburst: is it a galaxy-wide phenomenon, or is it one isolated to the disk (which might be the case for a normal spiral) or conversely to the center (if a merger has taken place, or if the original galaxy was an elliptical).

Because of the poor spatial resolution obtainable for the distant galaxies, little is known about the spatial distribution of the starbursts in the distant galaxies, though results from the HST as well as ground based sites with excellent seeing are now becoming available. Couch et al. (1994) report little spatial variation in V-I colors (rest B and R) in galaxies in AC114 observed with HST, indicating that the star formation was widespread in their blue galaxies, which for the most part are disk-dominated systems. Lavery & Henry (1994) likewise report no color variations in galaxies in three clusters at z=0.2, observed in $0.7''$ seeing on Mauna Kea. Franx (1993) took spectra of an E+A galaxy at z=0.18 which he confirms as a disk galaxy with a young component whose light distribution is the same as the old component. Oegerle et al. (1991) report a strong color gradient in a field E+A at z=0.09, though most of the luminous mass of the galaxy lies within the post-starburst area. Finally, Dressler et al. (1994), Couch et al. (1994), and Wirth et al. (1994) report that an unusually large fraction of the galaxies in Butcher-Oemler clusters are spirals, many of which are interacting. This piece of information tends to argue that the enhanced star formation is a galaxy-wide phenomenon.

The recent discovery of a number of post-starburst galaxies in the Coma cluster, most of them located to the SW of the cluster center, provides us with the opportunity to study the spatial distribution of the enhanced star formation with even greater resolution than the studies cited. To that end, we have observed four of the Coma galaxies with a long-slit spectrograph. Those data, along with similar data taken of two nearby field E+A galaxies (NGC 3156 and IC 2035), and data on Coma emission line galaxies, including one that is presently undergoing a starburst, are presented in this paper. The spectra provide us with the run of absorption line strength with radius, as well as kinematical information bearing on the type of dynamical system, i.e., whether pressure-supported or rotationally-supported. Direct CCD images were also taken to provide color maps to bolster the spectroscopic evidence for the distribution of the A-star components. These were also used to measure light profiles, which tell of the existence of disks in all of the cases, thus clearing up the question of the structure of the galaxies, which morphological data had been equivocal about (Caldwell et al. 1993).

While the main focus of this paper is to discuss the spatial extent of the post-starburst activity in the Coma cluster early-type galaxies, the data on the starburst galaxy proves to be interesting as well, with disturbed kinematics found in its associated ionized gas which may provide clues to the cause of the starburst phenomenon. Finally, the emission spectra of several other Coma cluster early-type galaxies are discussed.

## 2. OBSERVATIONS

### 2.1. Spectroscopy

Caldwell et al. (1993) identified eight moderate to strong cases of post-starburst galaxies in the Coma cluster. All but one of these lie in a region 0.5 Mpc to the SW of the center. We have obtained long-slit spectra of four of these galaxies with the MMT. Two galaxies which are believed to be field analogs of post-starburst galaxies were also observed, one with the MMT and another with the CTIO 1.5m telescope. We also report on the spectra of three currently active galaxies in Coma, two present day starbursts, the other a Seyfert galaxy. Table 1 lists the galaxies we observed, along with the pertinent observational data. The slit angle for D99 was chosen to include a neighboring irregular, D100. For D112, the major axis rotates through $30°$ between radii of 3 and $8''$; the position angle of $38°$ for the slit then refers to the major axis of the inner part of the galaxy. The pixel size in the spatial direction was $0.32''$ pixel$^{-1}$ for the MMT data, except for NGC 4853 in which case the pixel size was twice that. The seeing values during the observations are listed in table 1; typically, structures greater than 0.6 kpc in extent could be resolved in the Coma galaxies (assuming a Coma distance of 86 Mpc).

The MMT spectrograph CCD frames were reduced in a manner typical for long slit spectra. After subtracting off a zero level and a bias frame, cosmic rays were identified and flagged so that the pixels involved could be excluded when the frames were combined. Specifically, pairs of frames were subtracted to form a frame in which cosmic rays appeared as very high or very low pixels, which could easily be identified automatically and flagged; the flags were then transferred to the original frames. An average frame was then formed from the 3 to 4 individual frames available, each resultant pixel being the weighted average of the all the unflagged values for that pixel. Those frames were divided by a normalized flat field frame, which served to remove small-scale response variations and the non-flat slit response, but which left the spectral response alone. Companion arc line frames were fit with two dimensional functions, the latter were then used to transform the galaxy exposures to linear (or log) wavelength scales. The sky spectrum was determined from the edges of the slit and was subtracted from the entire frame. For the measurement of the spectral indices, the spectra were deredshifted using the redshift found for the nucleus of each galaxy. Since we planned on modeling the stellar populations of the nuclear spectra of the galaxies, we flux-calibrated the nuclear spectra using standard stars observed with the same instrumental setup.

## 2.2. Imaging

To aid in understanding the variation of spectral properties as a function of radius, we also obtained multicolor direct CCD frames of most of the galaxies. Various telescopes were also used for this part of the project, but all exposures were taken with B and R filters; see table 2.

These data were reduced as usual, by bias subtraction and flatfield division. Cosmic rays were either removed by hand or by filtering while combining multiple exposures. Accurate sky levels were determined by running an ellipse fitting program with a first estimate to the sky. A correction to the estimate was obtained by averaging the isophotes found by the program at very large radii; i.e., radii where the galaxy clearly has no light. The program was run again with this new estimate to the sky to find the light profile of the galaxy, along with the run of axis ratio and position angle as a function of radius. None of the observations were taken during photometric weather, so we only give relative surface brightnesses. Seeing values are listed in table 2, and typically correspond to 0.8 kpc for galaxies at the distance of Coma. We note that the resolution is thus similar or better than the resolution obtained with the HST PC data for galaxies in clusters at redshifts of 0.4 or greater.

We made color maps of the galaxies by subtracting the derived sky values from the frames, shifting the B and R images to a common center, and matching the point-spread functions of the frames. The frames were then simply divided. The run of color with radius was found by averaging the color data along the isophotal ellipses found on the R frames.

## 3. ANALYSIS OF POST-STARBURST GALAXIES

### 3.1. Nuclear Spectra

As a reminder of the unusual nature of these galaxies, the global spectra (the unweighted sum of the light along the spectrograph slit) are shown in Fig. 1. The characteristic strong Balmer absorption lines, prominent Ca II K line and G band, and little if any emission are clearly evident. The global spectrum of IC 2035 (an isolated S0) is all that we have, thus the galaxy will not be taken up again until the section on colors. We now proceed to discuss whether these post-starburst characteristics continue outside of the nuclei in NGC 4853, D94, D99, D112, and NGC 3156, i.e., those observed with the MMT. Fig. 2 shows the nuclear spectra (averaged over 2.5″) compared to the extra-nuclear spectra (between 2.5 and 5″) for the galaxies observed with the MMT.

We shall simplify our terms somewhat for this discussion by calling the young component the "A" component, and the older component, the "E" component. These components were measured at various distances from the galaxy nuclei along the slits in two distinct ways, first by measuring equivalent widths of several important absorption lines, and second by fitting the spectral features simultaneously with a K star template and a template representing a young star population spectrum.

## 3.2. Radial Variations in Line Strengths and Colors

### 3.2.1. Equivalent widths

A program was written to extract equivalent widths from the spectra for the Balmer absorption lines H8($\lambda$3889), H$\delta$, H$\gamma$, and H$\beta$ as well as the K line of Ca II ($\lambda$3969). The Balmer absorption lines will of course indicate the presence of young stars (especially useful is the H8 line, which is totally absent in purely old populations) while the K line will be used as an indicator of the E component. If the A component decreases with respect to the E component, the Balmer equivalent widths should decrease, and vice-versa. The continuum windows used for each equivalent width measurement are listed in table 3. These were expanded somewhat for the lower resolution data on NGC 4853, so the data sets are not fully comparable, but are consistent enough for the measurement of gradients. The exact placement of the windows is not critical since we are only interested in differential measurements within one galaxy, thus the small offsets in the values of the equivalent widths among the galaxies (particularly the K line) are unimportant for this discussion. To insure adequate spectral signal-to-noise, co-addition of areas was performed at large distances from the galaxy centers. Typically, sensible measurements could be made out to 8-10″ in the Coma galaxies (corresponding to 3.3-4.2 kpc).

To improve the signal-to-noise in the Balmer measurements, all four Balmer lines were averaged for all galaxies except NGC 4853. For the latter case, H$\gamma$ was left out of the co-addition because it was difficult to find a suitable continuum for that line, due to the lower resolution data and the nearby G-band. We further note that the average of the four lines is typically very close to the average of just the H$\delta$, H$\gamma$, and H$\beta$ lines, which is all that could be measured for a purely old population. Doing so for a spectrum which consists of the mean spectra of 70 normal early-type galaxies in the central region of Coma results in a value of 2.9Å. Fig. 3 shows the combined Balmer equivalent widths and the K-line widths as a function of radius along the slit. Errors were calculated by Monte Carlo techniques for the nuclear point and the outermost points only; the errors for the intermediate points should be in between the values for the extrema. We now briefly discuss the individual cases.

D94 shows no measurable change in the relative strength of the K and A components over $16''$, $8''$ in radius, or 3.3 kpc. Both D112 and D99 have detectable decreases in the strength of the A component with radius. The combined Balmer equivalent widths have dropped by about 25-40% from the nuclear values at a radius of $7''$ (2.9 kpc), and are thus very close to the 2.9Å value found for a purely old population. NGC 4853 has a change in relative strength of the two components, but it is in the sense that the Balmer lines are weaker in the center.

The field galaxy NGC 3156 has a very strong concentration of the A component near the center, though the A component is still strong at the last measurable point. This galaxy is 4.5 times closer to us than Coma, thus the increased spatial resolution has perhaps allowed us to see a gradient in population whose strength would be washed out by seeing in the Coma galaxies, such as D99 and D112 which have but modest gradients.

The main result of this work is that the A component appears widely distributed in these galaxies, sometimes stronger, sometimes weaker in the nucleus relative to the underlying old population. We now use a different method to measure the same spectra.

### 3.2.2. Stellar template fits

The galaxy spectra were run through a fourier fitting program which solved for the presence of two templates at once, one a K giant, the other a synthetic spectrum of a 1.5 Gyr population created using Bruzual & Charlot's (1995) evolutionary synthesis models (see §3.4). In the standard fourier fitting technique (cf. Sargent et al. 1977), a single template is used, from which the radial velocity and velocity dispersion of a galaxy spectrum are determined, as well as a basic line strength index. However, Franx (1993) simultaneously fits two template stars in the case where two very different populations contribute to the integrated light of the galaxy. In this way, the relative contributions of both the E and the A components to the galaxy absorption line spectrum are found, in addition to the usual kinematic information.

The results of the line strength determinations are shown graphically in Fig. 4. The left hand panel shows the line strength index as a function of radius for each component, while the right hand panel shows the run of flux contribution in the spectra with radius. Gradients in the relative contributions of the two components will show up as changes in the relative fluxes; overall offsets in the zero points between the two components are unimportant. Once again D94 is shown to have little change in the relative content of the E and A components as a function of radius, in accord with what was found above. D112 and D99 are again shown to have a measurable falloff in the strength of the A component with increasing radius. NGC 4853 has a falloff in the A component at the nucleus, as seen above. At first we thought this might be due to the presence of nuclear emission, which would fill in the Balmer absorption lines and decrease the equivalent widths there, and in fact there is nuclear emission as discussed below. However, the color maps also shown below confirm that the galaxy gets redder in the center. Thus we conclude that the young population is less prominent in the nucleus than at larger radii. The field galaxy NGC 3156 is seen in the figure to have a stronger A component in the central 2-3″ (0.19-0.28 kpc) than outward of that radius, but only a small gradient in the outer parts out to 20″ (1.9 kpc), more or less in confirmation of the individual Balmer line measurements.

### 3.2.3. B−R color profiles

B and R light profiles and B−R color maps were obtained from the CCD data for D45, D94, D99, D112, NGC 4853, IC 2035, and NGC 3156. The light profiles themselves are discussed below. The B−R profiles are plotted in Fig. 5; the units are in magnitudes with an arbitrary zero point. The effects of one sigma errors in the mean sky level (the dominant error at large radii) on the colors are shown as continuous lines above and below the colors derived from the adopted sky levels. The color data generally extends in radius out to 1.5 times that of the spectral data.

NGC 4853 is seen to have a large gradient such that the galaxy becomes bluer with radius, thus confirming the spectral gradient. The lack of any gradient in the spectrum of D94 is seen in its color profile, too. The gradient in the Balmer line strength of D112 is not well reflected in its color profile, in which no significant color change is noted over 10″ (4 kpc). D99 has a small color gradient, about 0.1 mag over 10″ , in the sense that the nucleus is bluer than the outer parts. This is in accord with the spectroscopic measurements. The strong A component gradient in NGC 3156 found in its spectrum is also reflected in the colors, which become 0.25 mag bluer from a radius of 1 kpc to the galaxy's center. IC 2035, for which we have no spatial information from the spectra, shows a blue nucleus, followed by a gradual reddening of the light with increasing radius. The starburst galaxy D45 has a central region that is somewhat bluer than the outer parts, as would be expected given the distribution of the starburst, as discussed below.

In summary, the color profiles confirm and extend the findings from the spectra of these galaxies: the starbursts were not located exclusively in the nuclei and instead were spatially extended over typically several kiloparsecs. While on the one hand we find the starbursts to have been spatially extended, they are on the other hand much more centrally concentrated than the star formation typically found in early-type spirals, where the star formation is found spread throughout the disk but avoids the bulge-dominated central region (Caldwell et al., 1991).

## 3.3. Disk Nature of the Galaxies

### 3.3.1. Light profiles

All of the light profiles are shown together in Fig. 6, with relative intensity plotted against the semi-major axis of the best fitting ellipses in kpc. The data for IC 2035 and NGC 3156 are in the B band, all the others are in R. The profiles are plotted out to a radius where the error in the surface brightness is 100% (0.75 mag).

Differences of opinion between Caldwell and Dressler regarding the morphology of some these galaxies were cited in Caldwell et al. (1993); the point of disagreement centering on whether a galaxy was an E or an S0. Clearly these galaxies are in fact all disk systems judging from the exponential nature of the profiles. Decompositions of the bulge and disks were attempted, and listed in table 2. NGC 4853, D112, and IC 2035 appear to have large bulge to disk ratios, but not unusually large. (Sparke et al. 1980 presented a profile of NGC 4853 as well, concluding the same for that galaxy). Therefore, despite the confusion concerning the morphological classifications for these galaxies, they must all be S0's except for D45, which is classed an Sa because of the evidence for current star formation.

### 3.3.2. Kinematics

An additional indicator of the disk-like nature of a galaxy is high rotation with respect to its internal velocity dispersion. The fourier fits done to the spectra to derive the relative contributions to the light of the two components also provide velocity information as a function of radius. The velocities for each component are shown in Fig. 7, uncorrected for inclination. The spectral data do not go out very far in radius, so the true maximum velocities are probably somewhat larger than what we have estimated, which are listed in table 4.

NGC 4853 and D94 show the clearest evidence for the rapid rotation expected for a disk galaxy, with maximum velocities of 200 and 100 km s$^{-1}$, respectively, occurring within 7″ (2.3 kpc). The rotation in NGC 3156 is found to be about 75 km s$^{-1}$, similar to the $80 \pm 8$ that Bender & Nieto (1990) found. However, little rotation is found for D99 and D112. The lack of velocity gradient in D99 is not too surprising as the galaxy is nearly circular on the sky (b/a = 0.8), but the same result for D112 requires a little explaining if we are to believe the galaxy is a disk. Upon re-examining the image of D112, we realized that the galaxy has a strong twist in the isophotes, with the position angle of the best fitting ellipse changing 30° over the radius range 3-8″. The spectrograph slit was aligned with the inner position angle, so if the kinematic major axis is actually indicated by the outer position angle, then we have measured the velocity 30° away from the line of nodes. Combined with the inclination effect (see table 4), the true rotation in the plane of the disk should be the measured rotation divided by 0.49. The measured rate of $\sim 52$ km s$^{-1}$ would thus correspond to 107 km s$^{-1}$ in the disk, which seems reasonable for the galaxy's luminosity as predicted by a Tully-Fisher relation (Schommer et al. 1993). We conclude that the rotation rates are all consistent with the galaxies being disk systems.

The resolution of the MMT spectra prevented us from measuring internal velocity dispersions from those data, but we also have available the spectra of the same galaxies reported in Caldwell et al. (1993), which have a resolution of about 3.8Å (corresponding to a velocity width of about $\sigma=100$ km s$^{-1}$ ). The same fourier fit program was employed, using a K giant star as the sole template. The dispersions and their errors are listed in table 4. As a check on accuracy, we measured dispersions for a number of galaxies in the Caldwell et al. (1993) sample that were tabulated in the Burstein et al. (1987) lists, and found differences smaller than 10 km s$^{-1}$ . Dispersions less than the spectral resolution are difficult to measure in low signal-to-noise spectra, so the dispersions we list that are smaller than 100 km s$^{-1}$ are only indicative. We have no spectrum of NGC 3156 from which to derive a dispersion, but note that Bender & Nieto (1990) measured a value of $70 \pm 5$ km s$^{-1}$ (leading to a V/$\sigma \sim 1$).

Dividing the maximum rotation rate by the dispersions (table 4), and further deriving the anisotropy parameter $(V/\sigma)^* = V/\sigma \cdot [\epsilon/(1-\epsilon)]^{-1/2}$, where $\epsilon$ is the ellipticity (Kormendy 1982), we find that D94, D112, and NGC 4853 all have $(V/\sigma )^*$ in excess of 1, indicating that the galaxies are rotationally supported. This information is then consistent with the finding from the light profiles that the galaxies are disk systems. D99 has a relatively low $(V/\sigma )^*$, but one still consistent with the photometry that suggests that it too is a disk system.

### 3.3.3. Comment

These Coma post-starburst galaxies all appear to be disk galaxies, rather than ellipticals. At the very least, we can conclude that the starbursts were not the result of mergers of more or less equal sized components, for such an event would have surely destroyed the disks (Barnes & Hernquist 1991). We point out again that of these galaxies only D99 has a companion (D100), and that companion has a velocity that differs from D99 by 4700 km s$^{-1}$ . It is more likely that non-disruptive interactions among galaxies or interactions between galaxies and the intra-cluster medium are responsible. New observations (Dressler et al., 1994, Couch et al., 1994) of distant Butcher-Oemler type cluster galaxies seem to show that a large number of the post-starburst galaxies are interacting systems, but also that some are undisturbed. Curiously, those undisturbed galaxies have elliptical galaxy morphologies, whereas the Coma galaxies (most of which are also undisturbed) are disk systems. Couch et al. (1994) conclude that at least two processes are responsible for the enhanced star formation in the distant cluster galaxies. It seems to be the case that strong interactions did not create the starburst galaxies seen in Coma.

## 3.4. Ages and Strengths of the Starbursts

In order to better place the Coma cluster post-starburst phenomenon in the context of the dynamical process (or processes) that is responsible for the starbursts, it clearly would be useful to obtain an accurate assessment of the duration of the starbursts and of the time elapsed since the termination of the starbursts. We have recently developed a method for age dating post-starburst early-type galaxies. The details of this method are presented in an accompanying paper (Leonardi & Rose 1995), so here we only briefly summarize the method. The main obstacle to deriving ages of post-starburst populations superposed on an old galaxy is to disentangle the age of the starburst from the strength of the burst relative to the old population. Specifically, a variety of combinations of burst strength and burst age can produce nearly indistinguishable broadband colors and Balmer line strengths. Hence there is a degeneracy in burst strength and age that is similar to the age-metallicity degeneracy that hinders work on old stellar populations. To break the degeneracy for burst populations we use the combination of a ratio in the strength of H$\delta$, relative to the neighboring Fe I$\lambda$4045 line, and a ratio in the strength of Ca II H + H$\epsilon$ versus Ca II K. The former index is referred to as the H$\delta$/Fe I index, while the latter index is referred to as Ca II. Both indices are based on ratios of residual central intensities in neighboring absorption lines. As a result, the indices decrease in value as the Balmer line contribution to the spectrum increases. While the H$\delta$/Fe I index changes fairly uniformly from K stars through A stars, the Ca II index undergoes relatively dramatic changes in the early-type stars. As a result, different burst strength and age scenarios follow different tracks in a Ca II versus H$\delta$/Fe I diagram. As is described in detail in Leonardi and Rose (1995), we have used G. Bruzual and S. Charlot's galaxy evolution models (Bruzual & Charlot 1995) to model the evolution of post-starburst galaxies in the latter diagram as a function of burst age and strength.

In Fig. 8 the three Coma cluster post-starburst galaxies, D99, D112, and D94, are plotted in the Ca II versus H$\delta$/Fe I diagram as a large filled triangle, square, and pentagon, respectively. The nuclear region and off-nuclear region spectra have also been analyzed. The nuclear and extranuclear data for D99, D112, and D94 are plotted as unfilled and skeletal large triangles, squares, and pentagons, respectively. Representative $\pm 1\sigma$ error bars are plotted for the nuclear data point for D112, where the error bars have been calculated based on the photon statistics in the galaxy and the sky and on the read noise and gain of the CCD. Error bars are approximately the same for D112 and D94, but are approximately a factor of two larger for D99. In addition, the error bars are about the same for the nuclear and extranuclear regions, and hence somewhat smaller for the global values represented by the filled symbols. Also plotted are the nearby post-starburst galaxy NGC 3156 (starred quadrangle) and the outer region of the Coma irregular galaxy D100 (starred pentangle). For reference, we reproduce a grid of points derived from linear combinations of post-starburst model spectra with the observed spectrum of a presumed old galaxy population. The old population point, which consists of the mean spectrum of 70 normal early-type galaxies in the central region of Coma, is represented by an open triangle, at the apex of the grid. Filled squares represent Bruzual and Charlot model spectra for a pure 0.3 Gyr-long starburst that is seen at various times after termination of the burst. For instance, the square designated by 0.0 represents the indices for the burst immediately upon its termination, while the square labeled as 1.0 represents the indices of the burst 1.0 Gyr after its termination. The linear combinations of burst and old population are represented by the dotted lines; the small crosses designate 10% increments in the balance of burst versus old population light, normalized at 4000

Å. As can be seen from Fig. 8, the post-starburst Coma cluster galaxies are typically ∼0.8-1.3 Gyr past the termination of their main starbursts, and the burst spectra contribute ∼60% of the light at 4000 Å. For the field post-starburst galaxy NGC 3156, the burst has about the same age, but is a much larger fraction of the total nuclear light. In the case of the Coma irregular, D100, we have the interesting situation that star formation is ongoing in the central 2 kpc in diameter (as evidenced by the emission line spectrum in this region), but outside of the central region star formation terminated approximately 0.5 Gyr ago.

Not plotted in Fig. 8 is the Coma post-starburst galaxy NGC 4853, for which a spectrum was obtained at lower (12 Å) resolution. Because the Ca II and Hδ/Fe I indices have some sensitivity to spectral resolution, to assess the burst age and strength in the case of NGC 4853 it was necessary to smooth the Bruzual and Charlot model spectra to 12 Å resolution and remeasure the spectral indices at this lower resolution. For brevity we do not show here the lower resolution analog to Fig. 8 with NGC 4853 plotted on it. However, we have found that the burst age is ∼0.9 Gyr and that the burst contributes 60% of the light at 4000 Å, or 53% of the light at 5000 Å. The latter figure can be compared to an analysis of NGC 4853 carried out by Sparke et al. (1980) in which they synthesized the spectrum of NGC 4853 with a combination of an A1V spectrum, an F0V spectrum, and the spectrum of the elliptical galaxy NGC 3379. They found an optimum fit in which NGC 3379 contributes 50% of the light at 5000 Å, and the A1V and F0V spectra contribute 15% and 35% of the light respectively. In effect, the combination of A1V and F0V spectra represent a somewhat cruder synthesis of a young population than the Bruzual and Charlot models used by us, but the overall result in terms of fractional contribution from the young population is the same in both cases.

The above modeling has been carried out under the basic assumption that the Coma cluster post-starburst galaxies experienced a short, strong burst of star formation in the recent past. Is it possible, instead, that these post-starburst galaxies are actually spirals which experienced a sudden cessation of their star formation activity, as opposed to a burst? To assess this "extinguished spiral" option, we have generated Bruzual and Charlot models in which a spiral galaxy star formation history is maintained for 15 Gyr and then sharply truncated. We have then examined this truncated population at various times after star formation ceased. Two representative star formation histories have been modelled. Evidence presented in Caldwell et al (1991) and in Kennicutt et al. (1994) indicates that late-type spiral and irregular galaxies have maintained a roughly constant star formation rate (SFR) over the Hubble time. On the other hand, early-type spirals have experienced substantially decreasing SFR's over their lifetimes. Consequently we ran one set of models with a constant SFR truncated after 15 Gyr and a second set with an exponentially declining SFR in which the SFR after 15 Gyr (i.e., at the moment of truncation) is 10% of the initial rate. The results of these models have been plotted in Fig. 8 as thin solid lines. The line marked "CON" represents the constant SFR models at various times after truncation, while the line marked "DEC" represents the evolutionary track of the exponentially declining SFR models after truncation. Thus these two sequences indicate the expected tracks of fading late-type and early-type spirals under the assumption that there is no significant underlying old bulge light. If there were significant bulge light present, then the integrated indices of the fading spirals would lie somewhere between the 100% truncated disk sequences portrayed in Fig. 8 and the old composite Coma apex point in the upper right of the diagram (see Leonardi & Rose 1995 for further explanation). The most important aspect of the location of the two sequences in Fig. 8 is that the Coma post-starburst galaxies tend to lie to the left of the sequences, i.e., in a region of the diagram that is outside of the zone occupied by the truncated spiral models. Even in the most extreme scenario that the Coma galaxies could be truncated late-type spirals with no underlying bulge light, as represented by the "CON" line, the Coma galaxies cannot be reproduced by the models. The discrepancy between galaxies and models is even greater if we adopt a truncated early-type spiral scenario. Thus we can defintiely exclude the possibility that the enhanced Balmer lines in the Coma cluster galaxies are due to the extinguished spiral hypothesis; instead, a recent starburst must have taken place to produce the observed behavior of the Coma galaxies in the Ca II versus $H\delta$/Fe I plane.

The $\sim$0.8-1.3 Gyr post-starburst timescales discussed above can be compared to the dynamical timescale of $\sim$2 Gyr proposed by Burns et al. (1994) for the passage of an initially bound substructure through the core of the cluster and out the other side. While the post-starburst timescales found by us are rather short compared to that indicated by Burns et al. (1994), it is not clear at what point in the passage of the substructure through the cluster core that starbursts would be provoked, especially since the mechanism driving the starbursts has yet to be identified. Furthermore, as is discussed in Leonardi & Rose (1995), the ages determined above are based entirely on solar abundance models. If the galaxies actually are somewhat metal-poor, their derived ages would be greater. Specifically, Leonardi & Rose estimate that if the galaxies have [Fe/H] = -0.4, then the age determinations would be $\sim$40% higher, which would make them closer in value to the Burns et al. dynamical age. We therefore cannot say whether there is a disagreement between the two time scales at the present stage of this investigation.

## 4. ANALYSIS OF CURRENT STARBURST AND EMISSION-LINE GALAXIES

### 4.1. Kinematics of the Ionized Gas in D45

Most of the abnormal spectrum galaxies in the interesting SW region of the Coma cluster are in a post-starburst evolutionary phase in which the main starburst event terminated $\sim$0.8-1.3 Gyr ago. However, the Sa galaxy D45 is *currently* undergoing a starburst, as evidenced by its nuclear $H\alpha$ equivalent width of 75 Å. This galaxy is thus particularly important in terms of providing clues to the origin of the starbursts in Coma cluster early-type galaxies. For instance, tidal interactions have been proposed as explanations for the Butcher-Oemler effect in distant clusters (e.g., Thompson 1988; Lavery & Henry 1988; Lavery et al. 1992; Dressler et al. 1994), so that perhaps by studying this one particular galaxy we can comment on such ideas.

The blue spectrum of D45 shown in Caldwell et al. (1993) exhibits both enhanced Balmer absorption lines and a strong, narrow-line emission spectrum. The emission line equivalent widths are well in excess of those expected for an Sa galaxy, and the excitation of the spectrum resembles that of an H II region. Our MMT red spectra confirm the basic H II region nature of the emission spectrum (in that the [NII]/H$\alpha$ ratio is characteristic of an H II region rather than an AGN, see Fig. 13), hence D45 is a currently star-forming galaxy. This information is more extensively described in Caldwell et al. (1993) (cf. Appendix A). We now turn to further interesting aspects of the MMT red spectra that may provide clues to the origin of its starburst. The red spectra consist of one high-quality spectrum obtained in March 1993 along the major axis of D45 (PA=110°) and four somewhat lower-quality spectra obtained in March 1995 at position angles 114°(i.e., also along the major axis), 0°, 50°, and 90°. The latter spectra suffer from wavelength shifts (of up to 50 km s$^{-1}$ ) during 2000 second exposures and also from temporal variations in spectral resolution. These effects can been seen both in the comparison arc spectra and in the night sky lines in the individual galaxy exposures.

First, the strong emission is confined to a radius of ∼3″ (or 1.3 kpc) of the nucleus, as can be seen in Fig. 9, where the ionized gas distribution is shown for the major axis position angle in the 1993 spectrum. At 3″ , the flux in H$\alpha$ has fallen to 5% of its central value. This basic pattern of high surface brightness emission in the central 3″ radius is repeated at the other three position angles, and is also confirmed in the major axis position angle for the 1995 data. Since the central high surface brightness emission coincides spatially with the inner bulge of D45, we hereafter refer to this emission as the "bulge" emission (refer to the light profile in Fig. 6), even though we show below that the "bulge" gas is probably distributed in a disk. Thus the currently starbursting region in D45, while clearly spatially extended, is confined to the bulge. In particular, it is very different in spatial distribution from the weak emission typically found in early-type spirals, where, as noted before, weak H II regions are spread uniformly throughout the disk and are not visible in the bulge-dominated central region.

Second, faint disk H$\alpha$ emission can be traced out to 23″ (9.6 kpc) from the nucleus on the east side of the slit (to the bottom of Fig. 9), This emission (which we will refer to as the "disk" emission, since it is spatially coincident with the low surface brightness disk of D45) is presumably produced by low-luminosity H II regions that are typical for an Sa galaxy (Caldwell et al. 1991). It is also barely discernible in the lower-quality 1995 major axis long-slit spectrum. Furthermore, no faint extended emission is seen on the west side of the slit, nor is it seen at any of the other three position angles. The asymmetric nature of the disk emission appears to be correlated with a similar asymmetry visible in the B and R images of D45. A grey scale representation of the R band image is displayed in Fig. 10, where it can be seen that there is extended low surface brightness structure on the eastern side of D45 that is not present on the western side.

Third, the bulge and disk emission reveal unusual relative kinematics. The bulge gas kinematics are well-defined by rotation curves at the four position angles covered in the 1995 MMT red spectra. The rotation curve data is summarized in Fig. 11, where it is seen that the bulge gas kinematics displays the characteristic signature of major axis rotation. This conclusion is further supported by the fact that the H$\alpha$ emission line widths correspond to a velocity dispersion of only $\sim$30 km s$^{-1}$, which is also characteristic of a fairly cold, rotating disk. The unusual kinematics become apparent in the high-quality 1993 major axis long-slit spectrum, which is shown in Fig. 12. The bulge gas rotation curve is strange, in that the gas west of the nucleus (as defined by the maximum in the continuum light) shows a velocity gradient, but that east of the bulge does not. Furthermore, the faint disk emission found *east* of the nucleus has a similar velocity to that of the gas *west* of the nucleus (cf. Fig. 12). The odd kinematics of the bulge gas with respect to the disk gas could mean that the bulge gas is counter-rotating with respect to the disk, or at least that the plane of the bulge gas is substantially different from that of the disk. The presence of a bar could cause some peculiar kinematics, though it would not easily explain the assymmetry of the gas distribution and kinematics. Assuming the velocity of the disk gas represents the "normal" reference frame, then it would appear that the origin of bulge gas is probably external to the galaxy, which would be an interesting piece of evidence in explaining the origin of the Coma cluster starbursts.

Finally, we note that the electron density of the ionized gas in the bulge of D45, as determined from the ratio [SII]$\lambda$6717/[SII]$\lambda$6731, is in the low density limit (i.e., $n_e \leq 10^2$ cm$^{-3}$) throughout the bulge region. The low density limit is typical of H II regions.

Clearly it would be interesting to determine the spatial distribution of the young star component, i.e., whether it covers a radius of several kpc as is the case for the other galaxies studied here (which would mean that the starburst was originally more extended than it is now), or whether the A star component is co-extensive with the present star formation as indicated by the ionized gas. To evaluate this requires the acquisition of long-slit spectra in the blue.

## 4.2. Line Ratios and AGN Signatures for D15, D16, D44, NGC 4853 and NGC 3156

The definition of an E+A spectrum as given by Gunn (1988) specifies that in addition to strong Balmer absorption lines, no emission lines (which could be evidence for ongoing star formation) should be detected. Thus a galaxy so described must truly be considered in a post-starburst phase. D94, D99, and D112 all have no detectable [OII]$\lambda$3727 to a limit of 0.5 Å, thus clearly meeting the most restrictive definition. Caldwell et al. (1993) found that some early type galaxies in the Coma cluster have strong Balmer absorption lines *and* emission lines. However, the emission lines are much too weak (relative to the Balmer absorption lines) to be consistent with an ongoing vigorous burst of star formation, and thus these galaxies as well must be basically characterized as in a post-starburst phase. In fact, the emission in some of the cases is not even thermal in origin, and hence is definitely not a signature of active star formation. We now continue this discussion with improved emission-line spectra for four other Coma SW region galaxies with emission, and comment on the emission line spectrum of the field galaxy NGC 3156.

Fig. 13a shows the sky-subtracted red MMT spectrum of D16. In addition to narrow-line [NII]$\lambda$6548,6584 and H$\alpha$ there is prominent broad-line H$\alpha$ emission. The spectrum of D16 is in fact very similar to the spectrum of M81 shown in Fig. 2 of Peimbert and Torres-Peimbert (1981), and, along with the high-excitation and He II$\lambda$4686 emission previously found in the blue spectrum, exhibits the main features of a Sy 1 spectrum. Both the broad-line and narrow-line emission in D16 is confined to the central $\sim$2″, i.e., the emission is not spatially resolved.

D44 is another Coma cluster galaxy with an emission spectrum characteristic of an active galactic nucleus accompanied by strong Balmer absorption (Fig. 13a). We have attempted to correct for the H$\alpha$ absorption accompanying the emission by subtracting off an appropriately scaled spectrum from off-nuclear regions that have no emission. The [NII]$\lambda$6584/H$\alpha$ line ratio so derived is about 2.0, somewhat lower than what Caldwell et al. (1993) had estimated. Still, D44 has an emission spectrum occupying a transition between LINER and Sy2 spectra. As in the case of D16, the emission lines in D44 are very centrally concentrated. An additional AGN spectrum in the SW region of Coma is present in D15, a Markarian object (Mrk 55), which, as is described in Appendix A of Caldwell et al. (1993), is the composite of an H II region spectrum and an AGN. Our MMT H$\alpha$ spectrum shows broad wings in the SII and NII lines (FWZI$\sim$1600 km s$^{-1}$ ; see Fig. 13b) within 2″ of the nucleus. Again the emission is very centrally concentrated, but is clearly spatially resolved, and exhibits a full amplitude rotation of $\sim$100 km s$^{-1}$. Hence three AGN spectra have been discovered in the SW region of the Coma cluster, which is remarkable given the extremely low general incidence of AGN spectra in the Coma cluster and in other clusters.

In contrast to the above three AGN spectra, NGC 4853 has an emission line spectrum of an H II region, with a [NII]$\lambda$6584/H$\alpha$ ratio of about 0.5 when one takes into account the H$\alpha$ absorption. The velocity dispersion of the lines is about 140 km s$^{-1}$, which, when allowance is made for the contribution of rotation to the line widths, is also consistent with the emission source being weak nuclear H II regions. The equivalent width of H$\alpha$ is only 7 Å, and the star formation rate implied by the H$\alpha$ luminosity is less than 1 M$_\odot$ yr$^{-1}$, thus even though the galaxy is ostensibly still forming stars, the burst phase is certainly over.

Finally, we consider the emission spectrum of the nearby post-starburst galaxy NGC 3156, which shows weak [OIII]$\lambda$5007 emission spatially extended over a region 12″ in diameter, as determined from the MMT blue spectrum. A spectrum in the red obtained with the FLWO 1.5m (Fig. 13a) reveals that the equivalent width of H$\alpha$ emission is approximately equal to that of the neighboring [NII] lines, a situation that is typical for LINER spectra. Thus the emission spectrum of NGC 3156 also is dominated by an AGN.

## 5.  SUMMARY

From the data in this paper, we conclude that the Coma cluster post-starburst galaxies are disk galaxies, but not simply spirals in which the disk star formation was extinguished. Instead, they had to have undergone a strong starburst that involved much more star formation than is typical for a spiral, and which occurred for only a short period of time. As well, the star formation is spatially extended in the galaxies, but more centrally concentrated than is the star formation found in early-type spirals (Caldwell et al. 1991). Our evidence for this is the long-slit spectra and images which show that the post-starburst signatures are prominent in the central region *and* are spatially extended. Further evidence from age dating the stellar populations shows that models consisting of 15 Gyr of constant star formation that suddenly terminate (i.e., an extinguished spiral) do not reproduce the observed spectral indices of the Coma post-starburst galaxies. Instead, we find that the Coma E+A galaxies discussed here had starbursts that ended 0.8 to 1.3 Gyr ago. The mass involved in the starbursts is roughly 10% of the total mass, thus the structures of the galaxies would not be greatly altered even if the starburst distributions were very different from that of the underlying galaxies.

Finally, it is of interest to compare our constraints on the recent star formation activity in early-type galaxies in Coma with the work on spirals that has been carried out by Bothun & Dressler (1986), and more recently by Gavazzi and collaborators (e.g., Gavazzi 1989; Gavazzi et al. 1995) and by Moss & Whittle (1993). These studies have demonstrated that spirals in the central 2° radius of the Coma cluster are on average strongly deficient in neutral hydrogen. On the other hand, a substantial fraction of them have enhanced current star formation rates. This is particularly true of the *early-type* spirals, for which $\sim$30% show enhanced star formation rates, as evidenced by strong H$\alpha$ emission (Moss & Whittle 1993) or radio continuum emission (Scodeggio & Gavazzi 1993). Moreover, the enhanced star formation has been demonstrated by Moss & Whittle (1993) to be centrally concentrated. There is a striking similarity here with the early-type galaxies that we have studied in this paper, i.e., all have experienced a burst of recent star formation that is spatially extended but quite centrally concentrated. Further work on the unusual star formation histories as well as the spatial distribution of these enhanced star-forming spirals in nearby clusters should help elucidate their relation to the post-starburst S0's. For now, it appears that whatever mechanism(s) is responsible for the starbursts occurring at the present epoch in Coma cluster galaxies is generating a similar kind of star formation activity in both E/S0 galaxies and in early-type spirals.

We wish to thank Mark Phillips and Mario Hamuy for obtaining the IC 2035 data, in exchange for some supernova observations to be named at a later date. Most of the data for this work was obtained with the MMT, which is operated by the Smithsonian Institution and the University of Arizona. This research was partially funded by National Science Foundation grant AST-9320723 to the University of North Carolina-Chapel Hill.

| Galaxy[a] | Tel.[b] | Date | Exp. | Res. | $\lambda\lambda$ | PA[c] | Seeing[d] |
|---|---|---|---|---|---|---|---|
| D94 | MMT | 1993 30 March | 10000s | 7.3Å | 3740–5600Å | 7 | 1.6″ |
| D99 | MMT | 1993 31 March | 7500s | 7.3Å | 3740–5600Å | 121 | 1.3″ |
| D112 | MMT | 1993 31 March | 10000s | 7.3Å | 3740–5600Å | 38 | 1.3″ |
| NGC 3156 | MMT | 1993 31 March | 2400s | 7.3Å | 3740–5600Å | 50 | 1.3″ |
| | | | | | | | |
| NGC 4853 | MMT | 1992 31 Jan | 1732s | 12Å | 3740–5900Å | 95 | |
| | | | | | | | |
| D16 | MMT | 1993 30 March | 2400s | 3.3Å | 6070–7020Å | 0 | 1.3″ |
| D44 | MMT | 1993 31 March | 2400s | 3.3Å | 6070–7020Å | 26 | 1.3″ |
| D45 | MMT | 1993 31 March | 2400s | 3.3Å | 6070–7020Å | 110 | 1.3″ |
| D45 | MMT | 1995 23 March | 4000s | 3.8Å | 5515–7708Å | 0 | 2.4″ |
| D45 | MMT | 1995 23 March | 2000s | 3.8Å | 5515–7708Å | 50 | 2.4″ |
| D45 | MMT | 1995 23 March | 2000s | 3.8Å | 5515–7708Å | 90 | 2.4″ |
| D45 | MMT | 1995 23 March | 4000s | 3.8Å | 5515–7708Å | 114 | 2.4″ |
| | | | | | | | |
| NGC 4853 | FLWO 1.5m | 1994 10 Jan | 1200s | 4.4Å | 3500–7400Å | 90 | |
| NGC 3156 | FLWO 1.5m | 1994 10 Jan | 1200s | 4.4Å | 3500–7400Å | 90 | |
| IC 2035 | CTIO 1.5m | 1992 7 August | 600s | 16Å | 3500–7700Å | 90 | |

Table 1: Log of Spectroscopic Observations

[a]The Dressler (1980a) or NGC numbers; for other aliases refer to Caldwell et al. (1993).

[b]The telescopes used for the spectra. FLWO 1.5m refers to the 1.5m telescope on Mount Hopkins.

[c]The position angle of the slit.

[d]The FWHM of standard stars taken during the same night

| Galaxy | Tel. | Exp. | Filters | Seeing | d/b |
|---|---|---|---|---|---|
| D45 | FLWO 1.2m | 900s | BR | 2.2″ | |
| D94 | KPNO 0.9m | 1800s | BR | 2.0″ | 10.1 |
| D99 | MDM 2.4m | 1800s | BR | 0.9″ | 4.6 |
| D112 | KPNO 0.9m | 1800s | BR | 2.0″ | 0.64 |
| NGC 3156 | FLWO 1.2m | 2700s | BR | 2.0″ | 3.3 |
| NGC 4853 | FLWO 1.2m | 1200s | BR | 1.6″ | 0.27 |
| IC 2035 | CTIO 0.9m | 600s | BR | 1.5″ | 1.0 |

Table 2: Log of Imaging Observations

| Line | Continuum 1 | Continuum 2 |
|------|-------------|-------------|
| H8 | 3863–3871Å | 3903–3915Å |
| H$\delta$ | 4080–4088Å | 4115–4125Å |
| H$\gamma$ | 4319–4323Å | 4362–4375Å |
| H$\beta$ | 4803–4826Å | 4895–4907Å |
| Ca II K | 3919–3925Å | 3944–3954Å |

Table 3: Line Index Definitions

| Galaxy | $\epsilon$ | i[a] | $V_{rot}$ (km s$^{-1}$) (observed) | $\sigma$ (km s$^{-1}$) | V/$\sigma$ | (V/$\sigma$)$^*$ |
|--------|-----------|------|-----------------------------------|------------------------|------------|------------------|
| D94 | 0.38 | 53 | 90 ± 10 | 72 ± 26 | 1.3 ± 0.5 | 1.7 ± 0.7 |
| D99 | 0.25 | 42 | 23 ± 10 | 78 ± 19 | 0.3 ± 0.2 | 0.5 ± 0.4 |
| D112 | 0.20 | 37 | 52 ± 10 | 73 ± 15 | 0.7 ± 0.2 | 1.4 ± 0.5 |
| NGC 4853 | 0.18 | 36 | 270 ± 20 | 135 ± 13 | 2.0 ± 0.2 | 4.3 ± 0.8 |
| NGC 3156 | | | | 75 ± 10 | | |

Table 4: Kinematic Data

---

[a]Inclinations derived from measured axis ratio and assuming intrinsic axis ratio of 0.2.

---

Fig. 1.— Global spectra of the E+A galaxies in this study. Spectra of NGC 4853, D94, D99, D112, and NGC 3156 are from the MMT. The spectra of NGC 4853 and IC 2035 have lower resolution than the other spectra (see table 1).

Fig. 2.— Nuclear and off-nuclear spectra of the four of the E+A galaxies. 7 pixels (2.1″ ) were summed along the slit at the nuclear positions and also at 7 pixels on both sides of the nuclei for each galaxy shown. The nucleus of D100 has strong emission lines; only the outer part which has a post-starburst spectrum is shown here.

Fig. 3.— Balmer (plotted as plus marks) and Ca II K (triangles) line equivalent widths as a function of radius for program galaxies. The $H\beta$ , $H\gamma$ , $H\delta$ , and H8 equivalent widths were averaged together to derive the Balmer values. Seeing diameters are indicated by horizontal error bars.

Fig. 4.— Line strength indices ($\gamma$ ) and relative fluxes for the E and A components as derived from a fourier fit to the spectra as a function of galactic radius. The E components are shown as open symbols, while the A components are shown as filled symbols. Seeing diameters are indicated by horizontal error bars.

Fig. 5.— B–R color profiles. The zero-point in the color is arbitrary. Radii of seeing disks are indicated below each profile.

Fig. 6.— Light profiles as derived from CCD images. The relative surface brightness in magnitudes is plotted as a function of radius, so that an exponential profile would appear as a straight line. The data for NGC 3156 and IC 2035 are from B CCD frames; the other data are from R frames. A vertical line at 3.3 kpc marks the radius inside which we have good enough spectral data to be able to comment on the relative proportions of the E and A components.

Fig. 7.— Stellar velocities of the E and A components as a function of radius, as derived by a simultaneous fit from a fourier fitting program. The E components are shown as open symbols, while the A components are shown as filled symbols.

Fig. 8.— Ca II versus H$\delta$/$Fe$ $I$ diagram for 0.3 Gyr burst models and for the Coma cluster post-starburst galaxies. The three Coma cluster post-starburst galaxies, D99, D112, and D94, are plotted as a large filled triangle, square, and pentagon, respectively; nuclear and extranuclear data for those are plotted as unfilled and skeletal large triangles, squares, and pentagons, respectively. Also plotted are the nearby post-starburst galaxy NGC 3156 (starred quadrangle) and the outer region of the Coma irregular galaxy D100 (starred pentangle). Curved, dashed lines are derived from linear combinations of post-starburst model spectra with the observed spectrum of a presumed old galaxy population, the latter represented by a large filled circle. Filled squares represent Bruzual & Charlot (1995) model spectra for a pure 0.3 Gyr-long starburst that is seen at the noted times after termination of the burst. The small crosses designate 10% increments in the balance of burst versus old population light, normalized at 4000 Å. Also plotted are thin solid lines, marked "CON" and "DEC" which represent the evolutionary tracks of truncated spirals seen at different times after termination of star formation. The tick marks at the bottom point of the lines represents the index values right after the truncation of star formation, while each successive tick mark denotes a time step of 0.5 Gyr. The "CON" line represents the constant SFR scenario while the "DEC" line represents the exponentially declining SFR described in the text.

Fig. 9.— Grey scale representation of the H$\alpha$ region of D45. The right hand portion has had the continuum emission removed so that the ionized gas can be more clearly seen. The strong starburst emission is confined to the area of strong continuum emission (the bulge of this Sa galaxy). Also note the peculiar kinematics of the ionized gas as discussed in the text, as well as the weak extended emission seen on only one side of the galaxy.

Fig. 10.— Grey scale representation of the R band image of D45. North is to the top, east to the left. Note the asymmetry of the low surface brightness emission in this galaxy.

Fig. 11.— Rotation curves from H$\alpha$ emission at four position angles in D45. All velocities are derived from the four 1995 MMT long-slit spectra.

Fig. 12.— Major axis kinematics for D45 derived from H$\alpha$ emission in the 1993 MMT long-slit spectrum.

Fig. 13.— The H$\alpha$ region of six galaxies in Coma. (a)The spectrum of D16 shows the weak, broad H$\alpha$ line. D44 is shown without correcting for the H$\alpha$ absorption present and also with a correction using the off-nuclear spectrum. D44 and NGC 3156 have emission line spectra of Sy 2 galaxies. (b) The H$\alpha$ region spectra of D45 and NGC 4853 are those of H II regions, while that of D15 is apparently a composite of an H II region and an AGN.

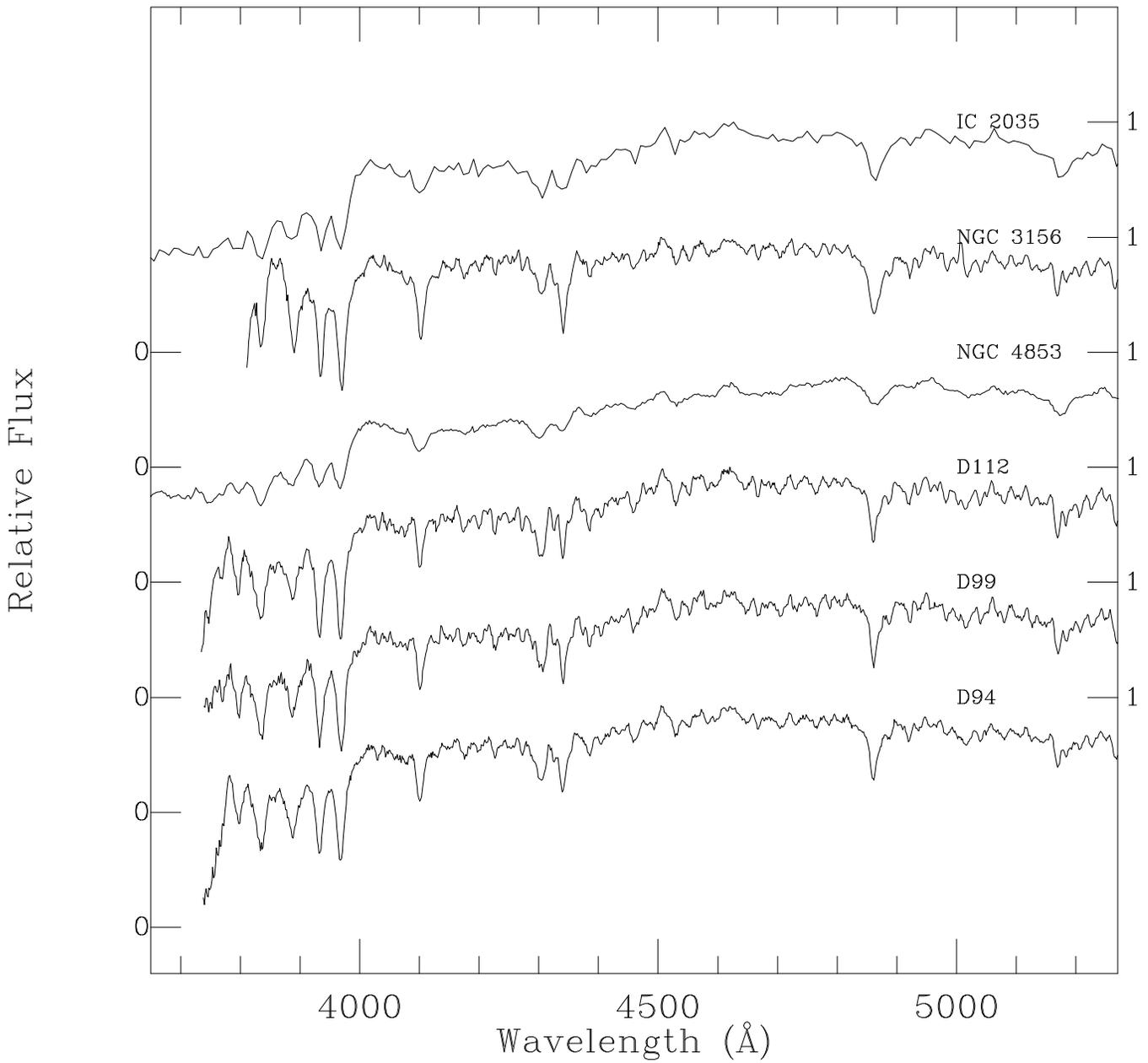

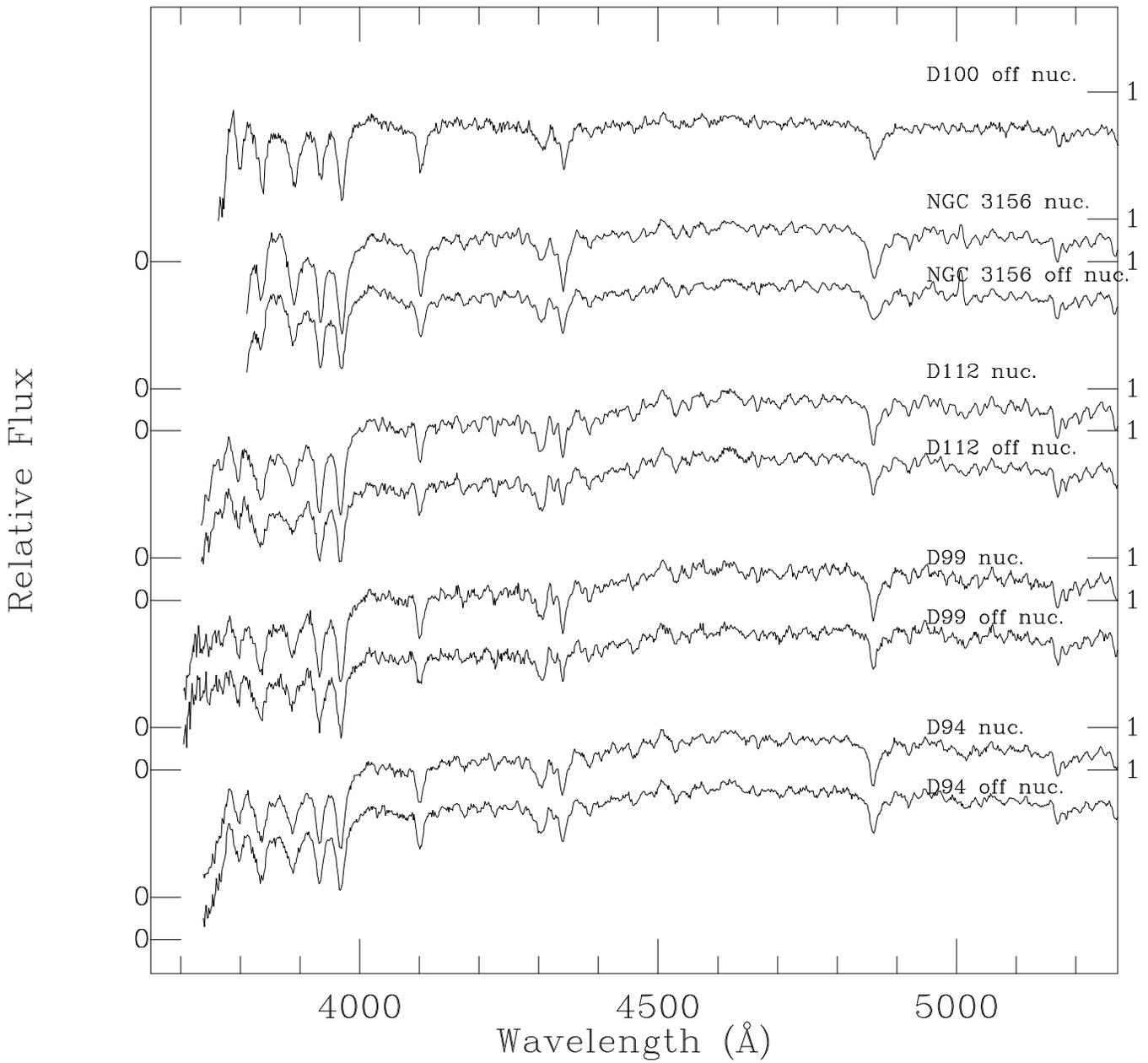

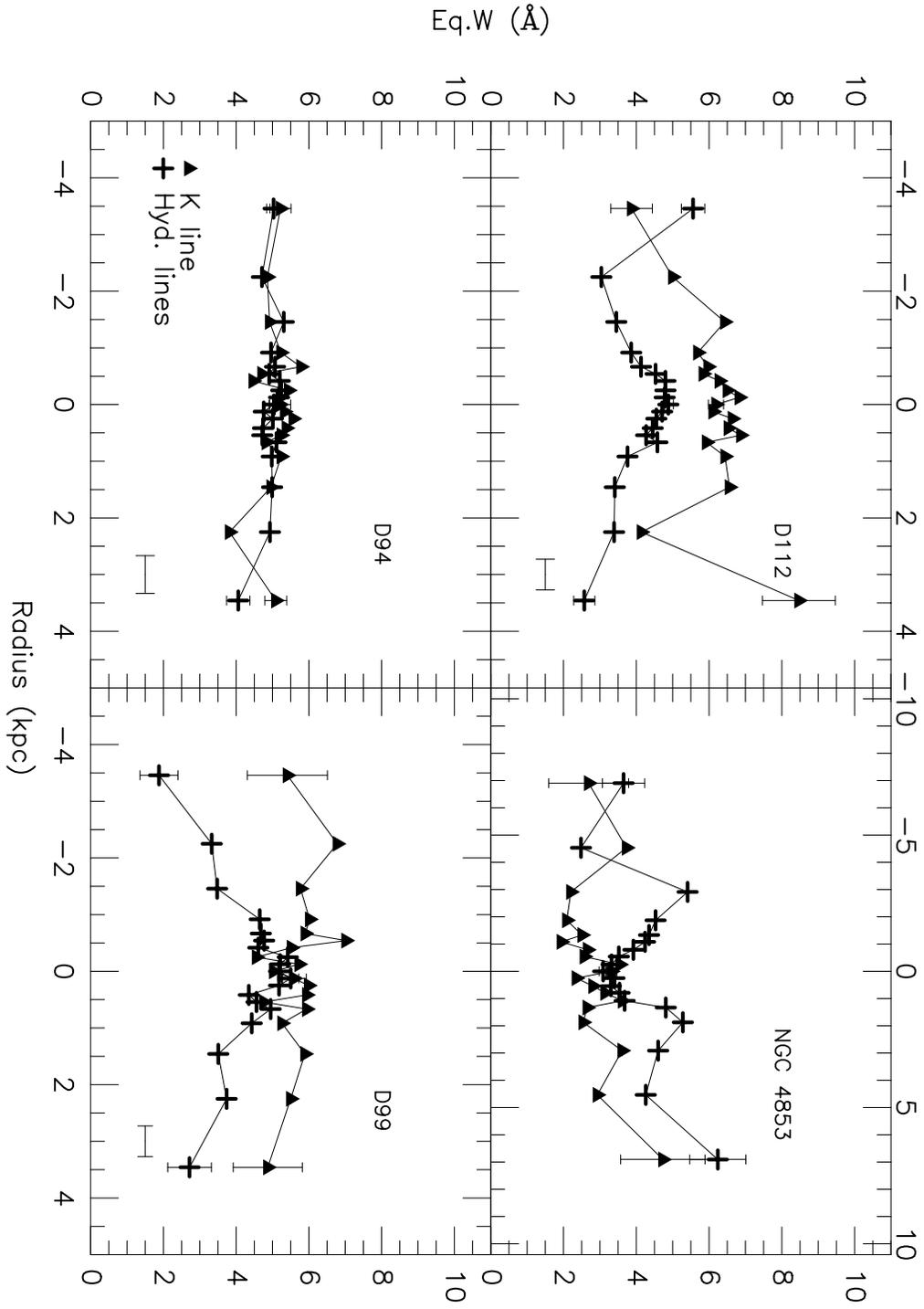

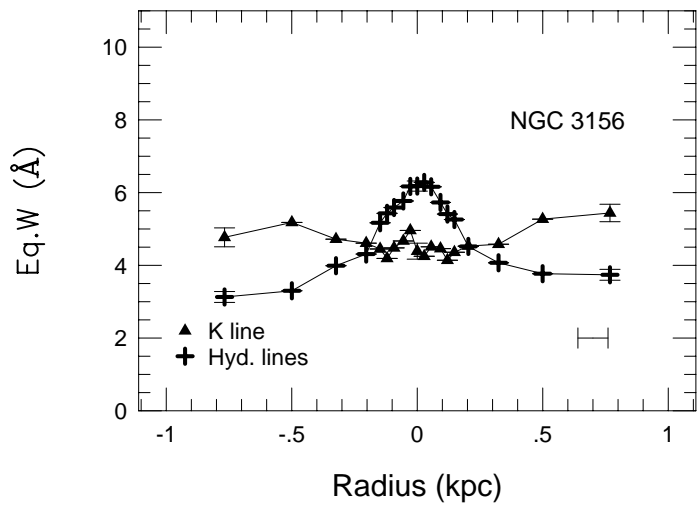

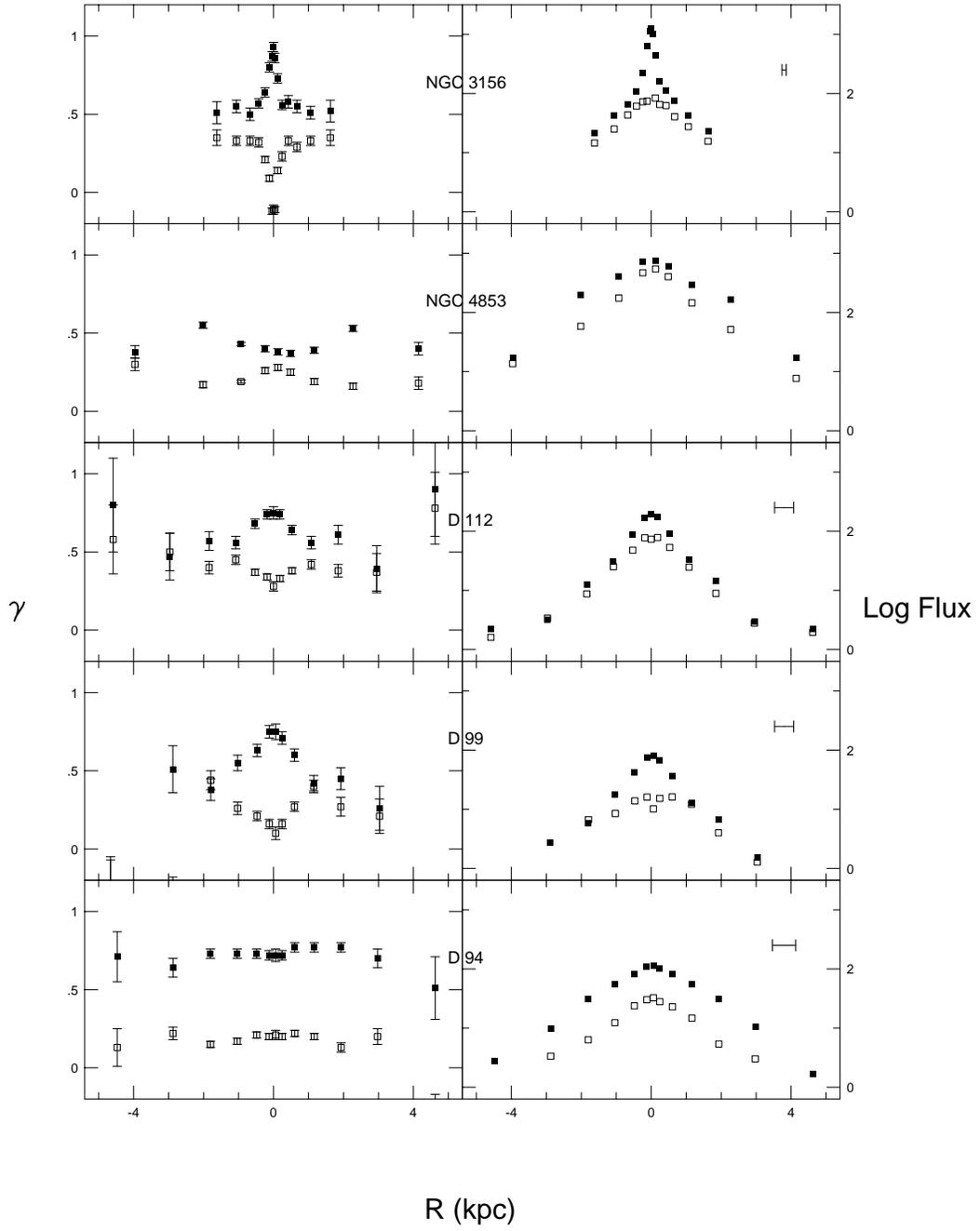

$\gamma$

Log Flux

R (kpc)

NGC 3156

NGC 4853

D 112

D 99

D 94

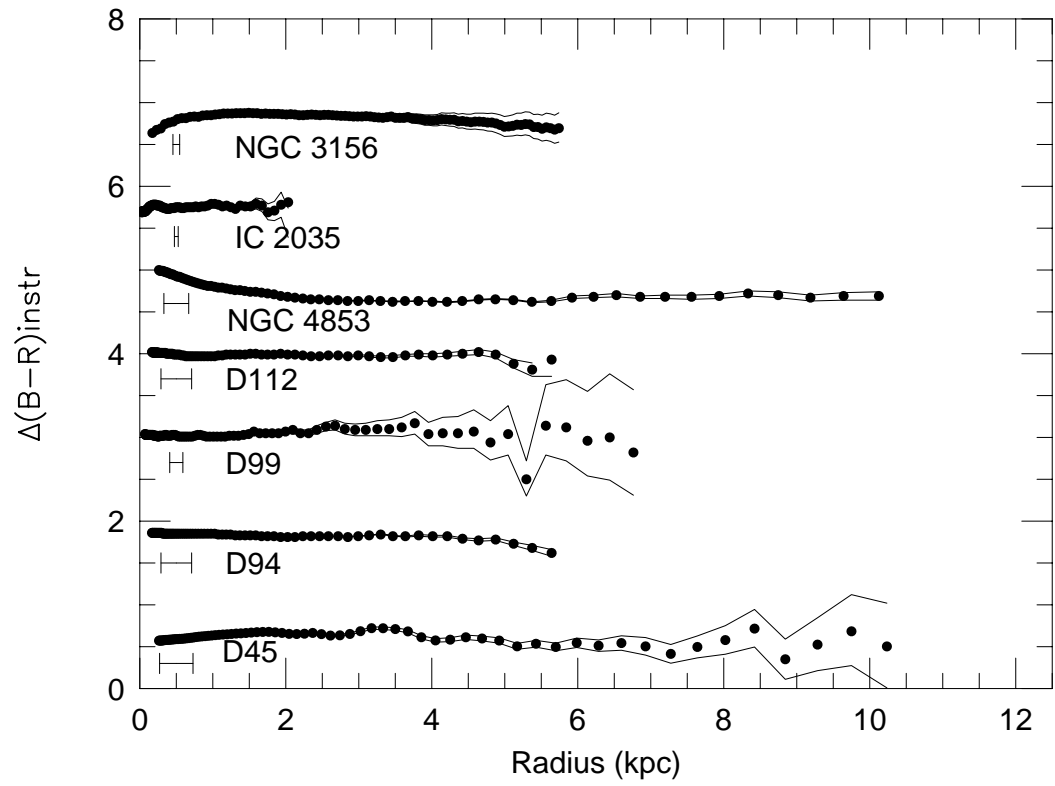

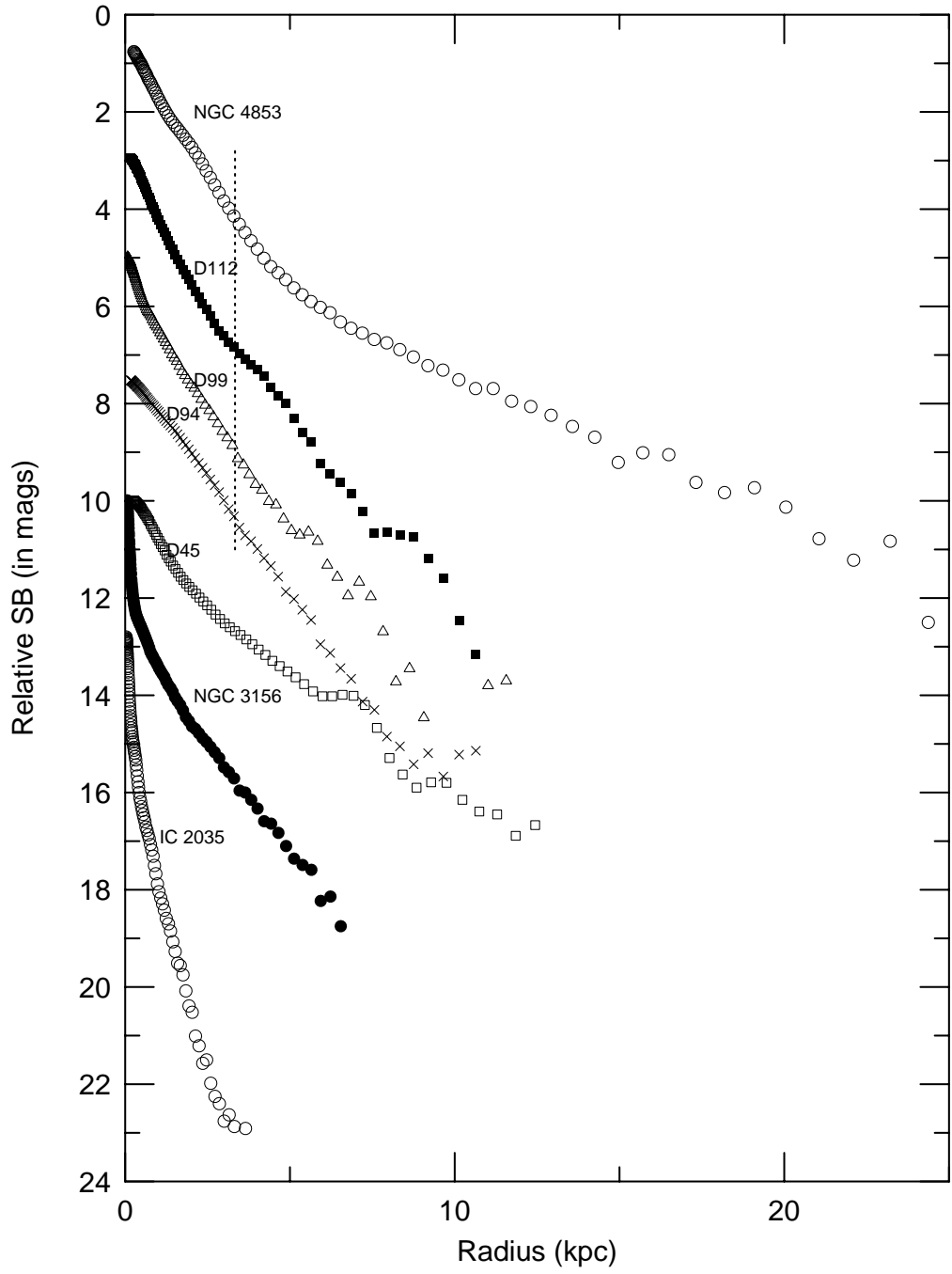

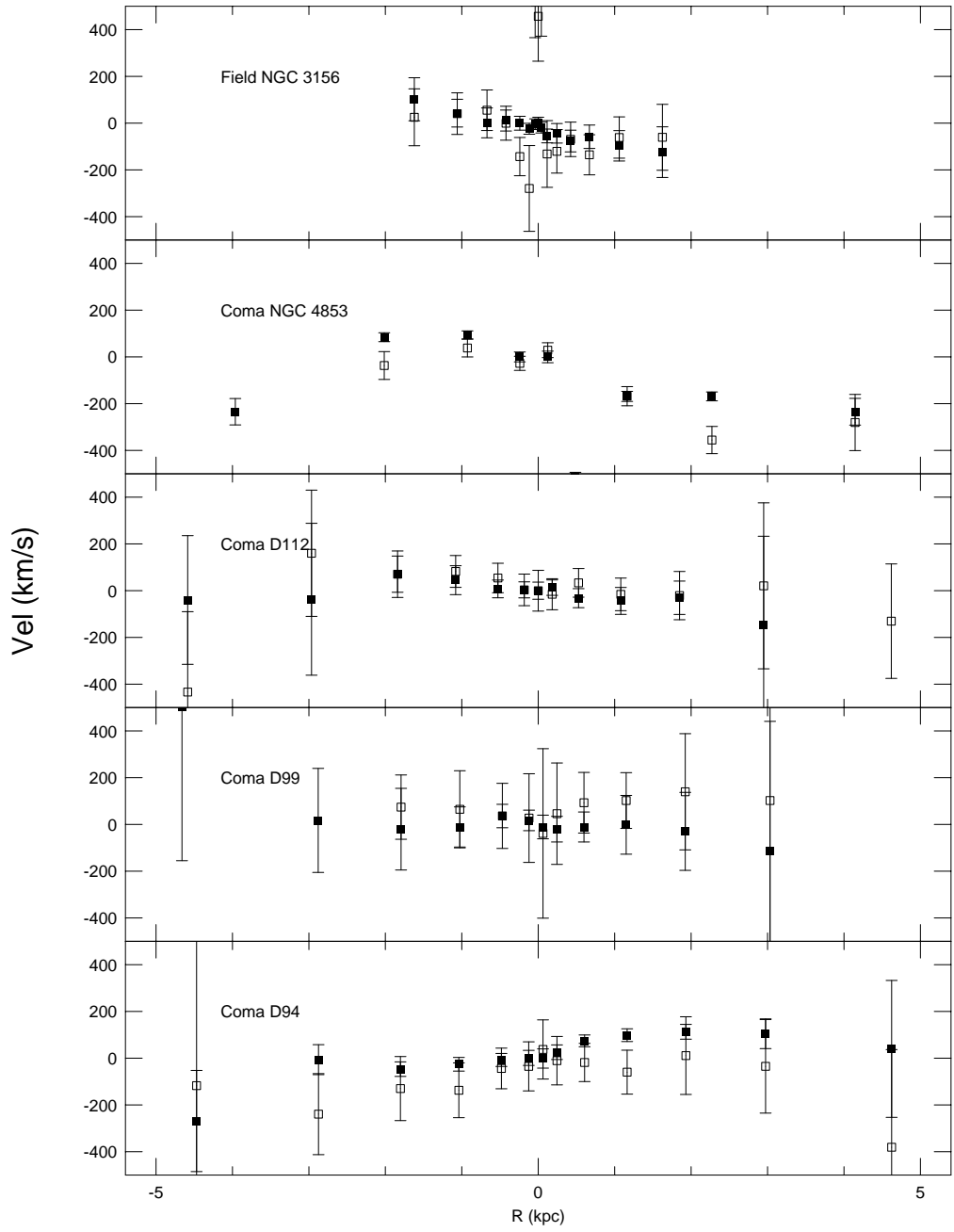

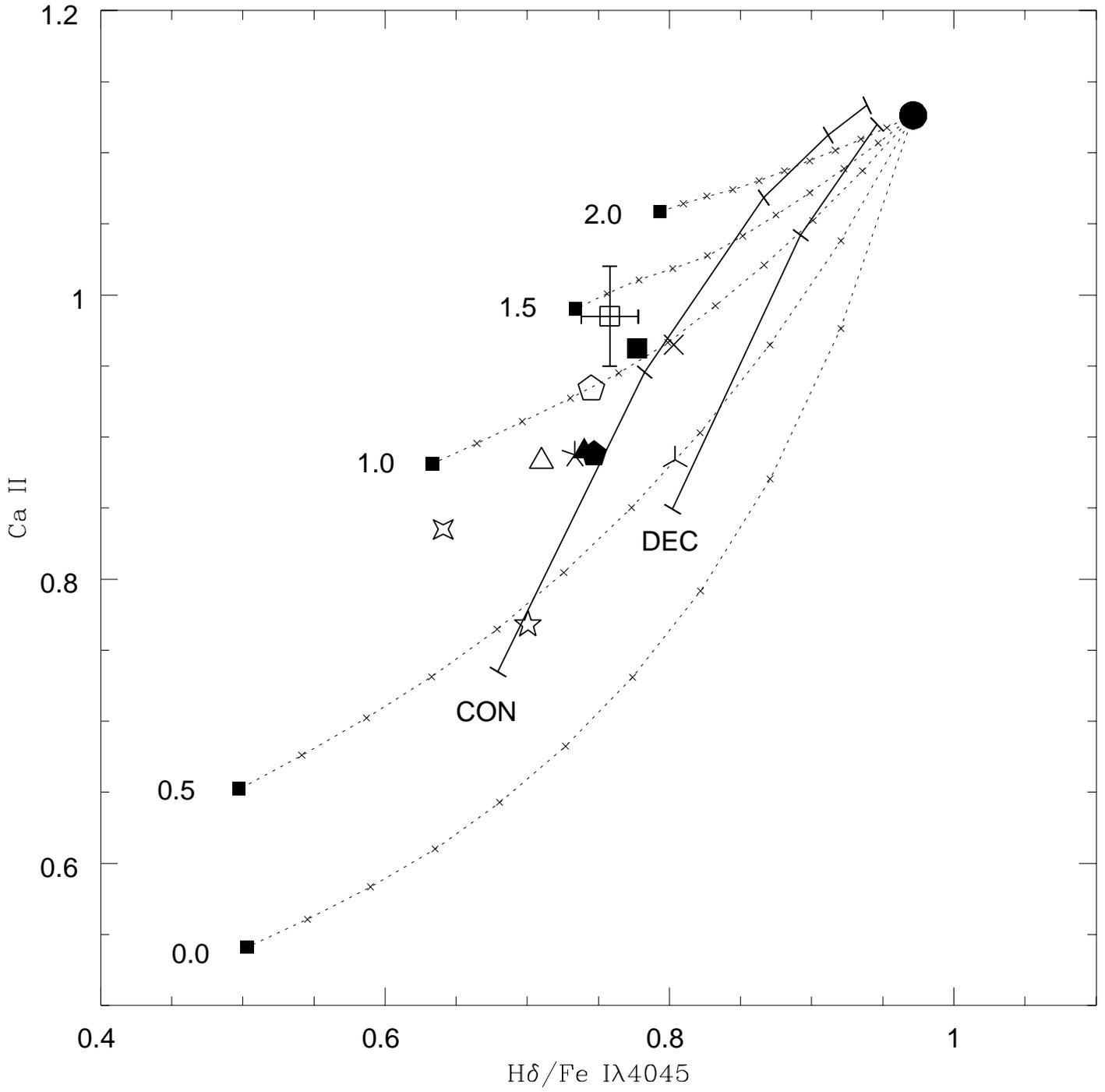

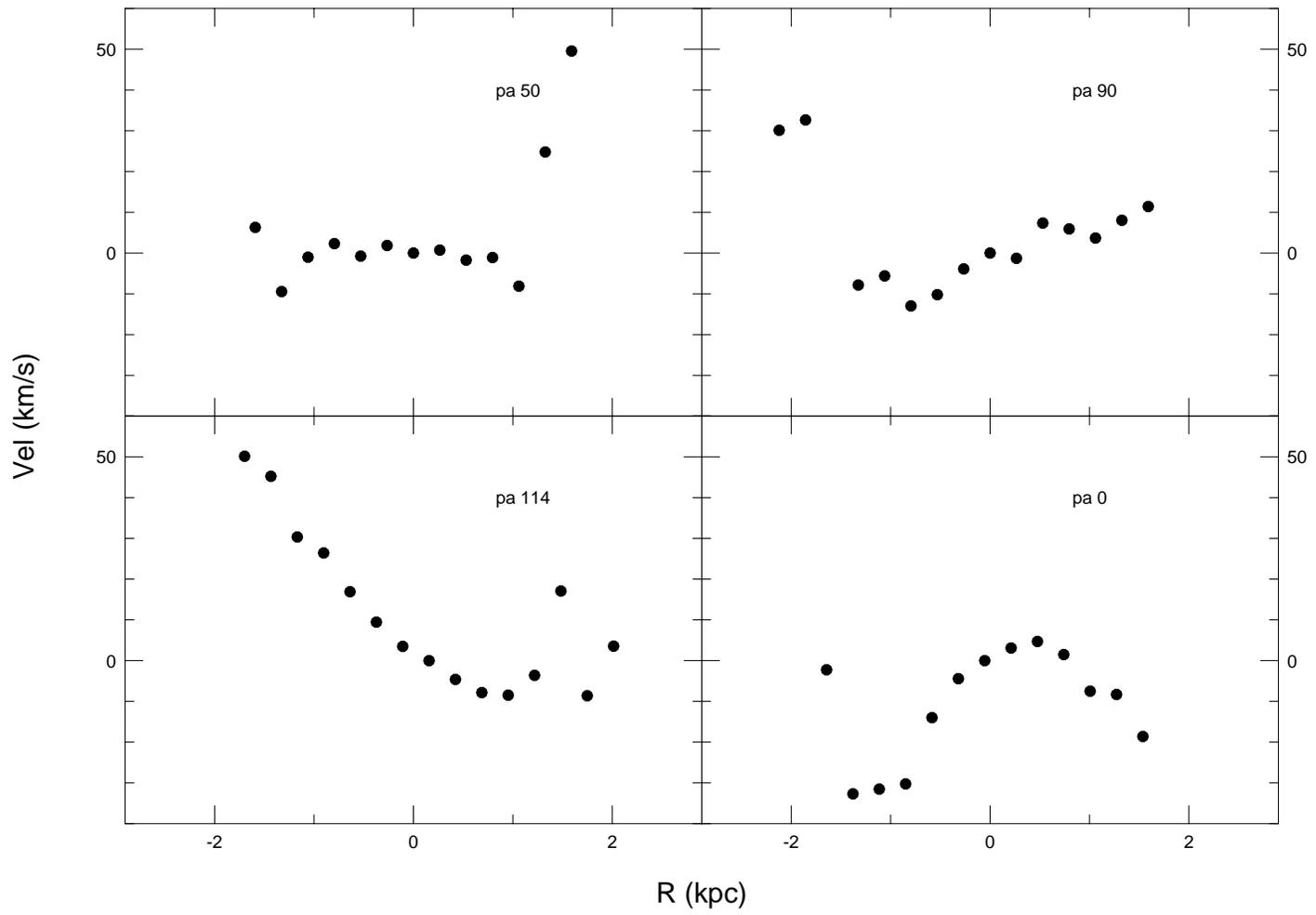

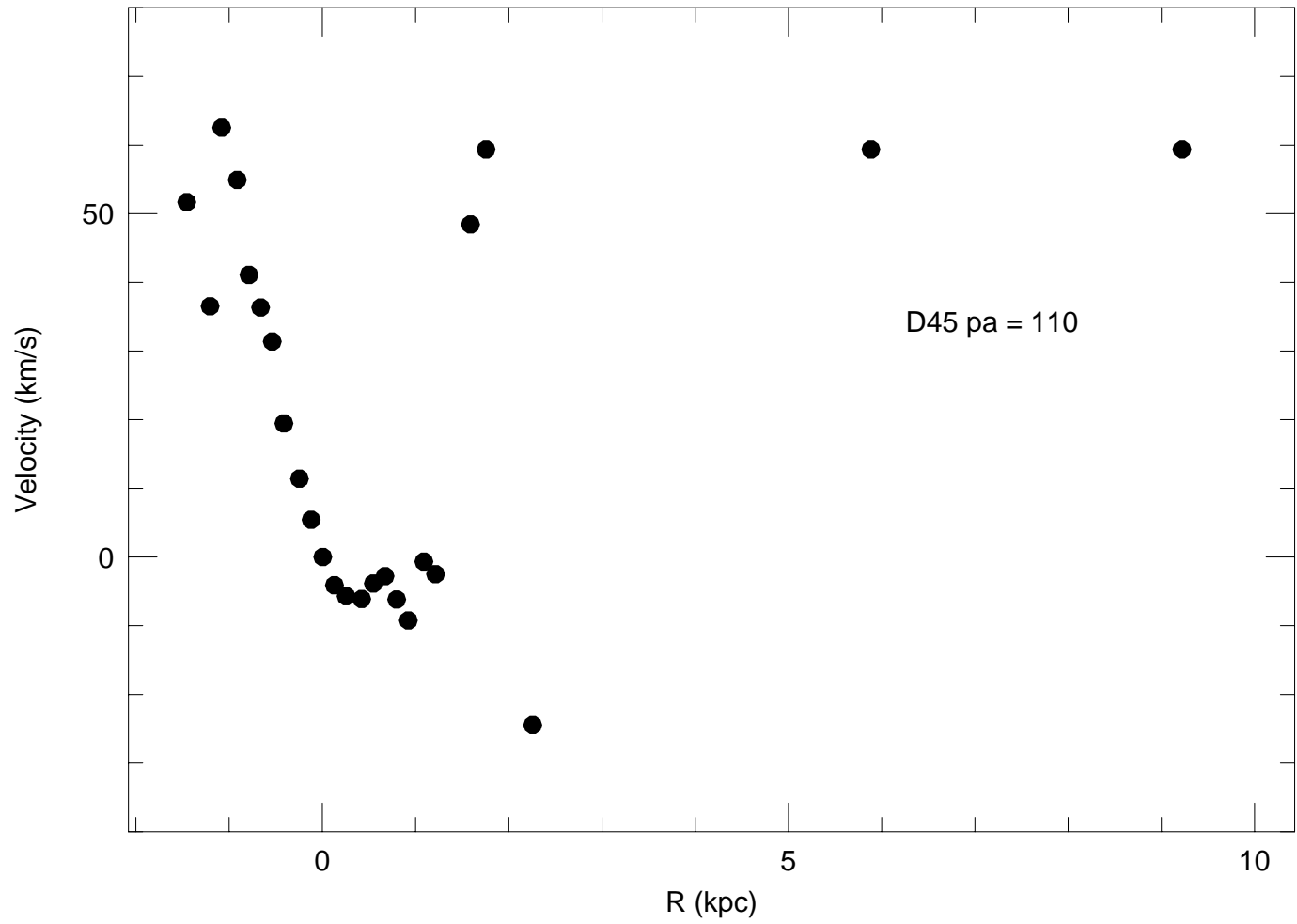

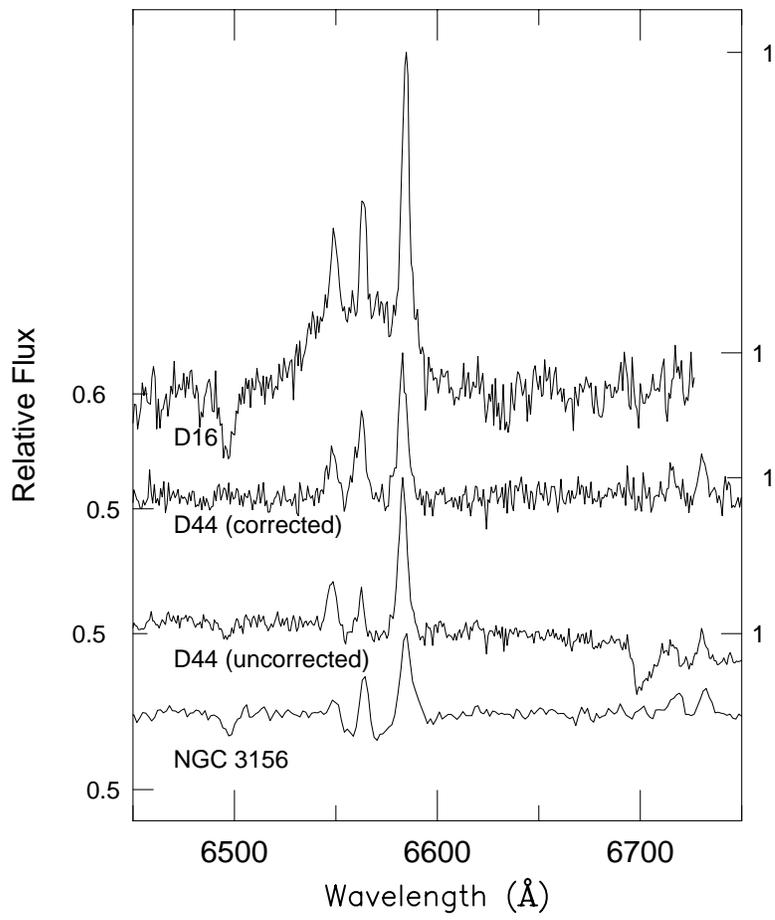

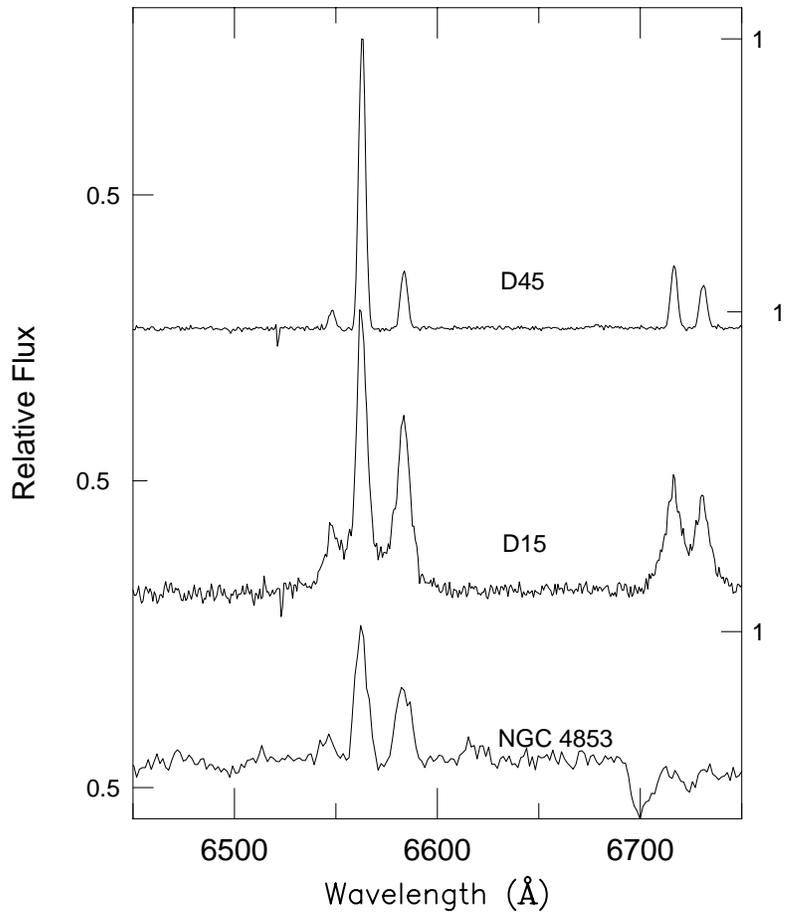